\documentclass[preprint,review,12pt]{elsarticle}

\usepackage{float}
\usepackage[table]{xcolor}
\usepackage{graphicx}
\graphicspath{ {./figs/} }
\usepackage{units} 
\usepackage{multirow}
\usepackage{amssymb}
\usepackage{pdfpages}
\usepackage{pdflscape}
\usepackage[english]{babel}
\usepackage[utf8]{inputenc}
\usepackage{booktabs}
\usepackage{ltablex}
\usepackage{lscape}
\usepackage{placeins}
\usepackage{hyperref} 

\usepackage{lineno}
\usepackage{amsmath}
\usepackage{dsfont} 
\usepackage{multicol}
\usepackage{relsize} 
\usepackage{import}
\usepackage{hyperref}
\usepackage{natbib}

\setcitestyle{authoryear,open={(},close={)}}

\journal{Spatial Statistics and accepted on 7 March 2020}

\begin{document}

\begin{frontmatter}


\title{Developments in statistical inference when assessing spatiotemporal disease clustering with the tau statistic}



\author[address1,address3]{Timothy M Pollington\corref{cor1}\href{https://orcid.org/0000-0002-9688-5960}{\includegraphics[height=8pt]{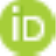}}}
\ead{timothy.pollington@gmail.com}
\cortext[cor1]{Corresponding author: MathSys CDT, University of Warwick CV4 7AL, UK}
\author[address2]{Michael J Tildesley\href{https://orcid.org/0000-0002-6875-7232}{\includegraphics[height=8pt]{figs/orcid_16x16.pdf}}} \author[address3]{T Déirdre Hollingsworth\corref{equal1}\href{https://orcid.org/0000-0001-5962-4238}{\includegraphics[height=8pt]{figs/orcid_16x16.pdf}}} \author[address4]{Lloyd AC Chapman\corref{equal1}\href{https://orcid.org/0000-0001-7727-7102}{\includegraphics[height=8pt]{figs/orcid_16x16.pdf}}}
\cortext[equal1]{Equal contributions from TDH \& LACC}

\address[address1]{MathSys CDT, University of Warwick, UK}
\address[address2]{Zeeman Institute (SBIDER), School of Life Sciences and Mathematics Institute, University of Warwick, UK}
\address[address3]{Big Data Institute, Li Ka Shing Centre for Health Information and Discovery, University of Oxford, UK}
\address[address4]{London School of Hygiene \& Tropical Medicine, UK}

\begin{abstract}
The \textit{tau statistic} $\tau$ uses geolocation and, usually, symptom onset time to assess global spatiotemporal clustering from epidemiological data. We test different factors that could affect graphical hypothesis tests of clustering or bias clustering range estimates based on the statistic, by comparison with a baseline analysis of an open access measles dataset. 

From re-analysing this data we find that the spatial bootstrap sampling method used to construct the confidence interval for the tau estimate and confidence interval (CI) type can bias clustering range estimates. We suggest that the bias-corrected and accelerated (BCa) CI is essential for asymmetric sample bootstrap distributions of tau estimates. 

We also find evidence against no spatiotemporal clustering, $p\textnormal{-value} \in [0,0\textnormal{\textperiodcentered}014]$ (global envelope test). We develop a tau-specific modification of the Loh \& Stein spatial bootstrap sampling method, which gives more precise bootstrapped tau estimates and a $20\%$ higher estimated clustering endpoint than previously published ($36$\textnormal{\textperiodcentered}$0m, 95\%$ BCa CI ($14$\textnormal{\textperiodcentered}$9$, 46\textnormal{\textperiodcentered}6), vs 30m) and an equivalent increase in the clustering area of elevated disease odds by 44\%. What appears a modest radial bias in the range estimate is more than doubled on the areal scale, which public health resources are proportional to. This difference could have important consequences for control.

Correct practice of hypothesis testing of no clustering and clustering range estimation of the tau statistic are illustrated in the Graphical abstract. We advocate proper implementation of this useful statistic, ultimately to reduce inaccuracies in control policy decisions made during disease clustering analysis.
\end{abstract}

\end{frontmatter}
\newpage

\thispagestyle{empty}
\begin{landscape}
\includepdf[pages=1,angle=90,pagecommand=\section*{Graphical abstract}]{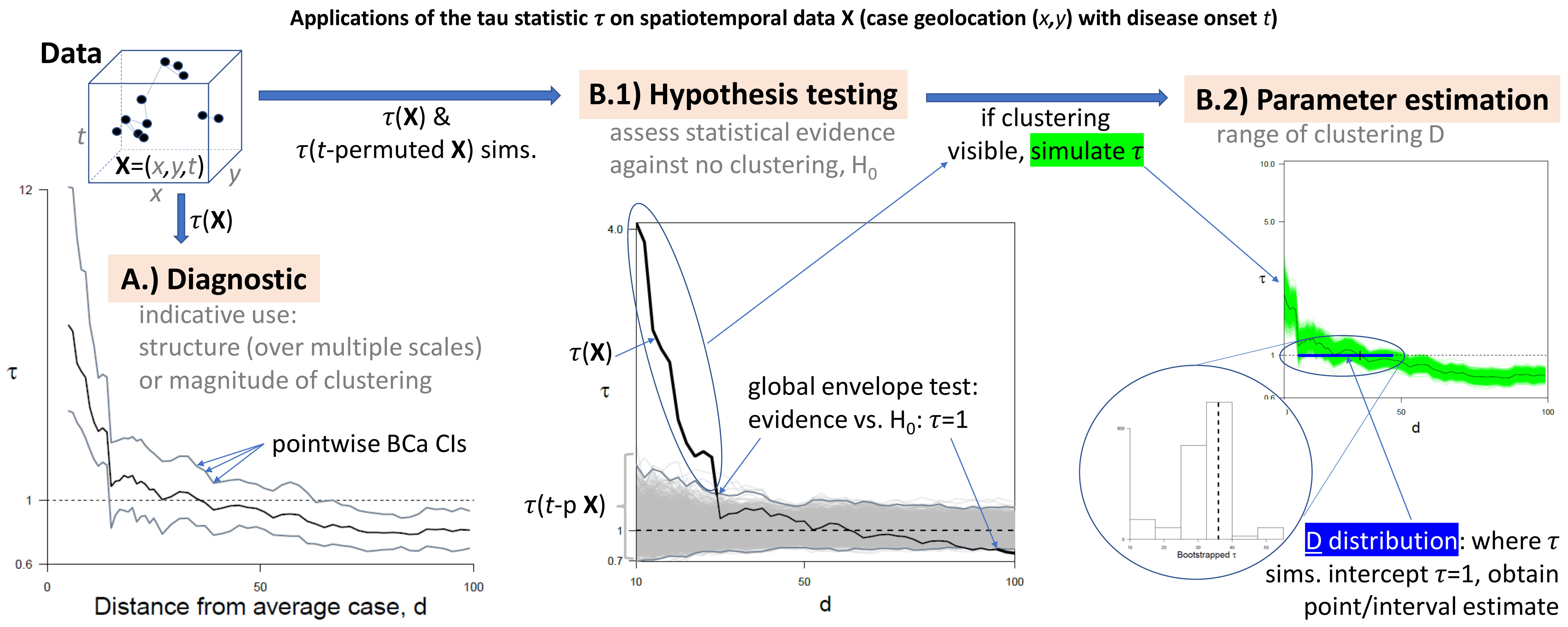}
\end{landscape}
\newpage
\newpage
\section{Introduction}
\label{S:Intro}

Assessing if \textit{spatiotemporal clustering} is present and measuring its magnitude and range is informative for epidemiologists working to control infectious diseases. The \textit{tau statistic} (\S\ref{S:thetaustatistic}) is more appropriate than most statistics for this task as it measures spatiotemporal rather than just spatial clustering, produces non-parametric estimates (without process assumptions) and, unlike the $K$ function \citep{Gabriel2009}, offers a relative magnitude in the difference of risk, rate or odds of disease (\S\ref{S:thetaustatisticodds}) versus the background level \citep{Lessler2016} \citep{Pollingtonreview}. The tau statistic herein should not be confused with `Kendall's tau statistic/rank correlation coefficient' \citep{bland2000introduction}. This study is motivated by a review of its use that found that its current implementation inflates type I errors (incorrectly rejecting a true null hypothesis) when testing for clustering, and may bias estimates of the range of clustering \citep{Pollingtonreview}. 

We investigate these aspects by analysing a well-studied open access measles dataset containing household geolocations and symptom onset times of cases (\S\ref{S:datasetandcomputmethods}). It represents a \textit{spatially discrete process} since infection is only recorded and can only occur at discrete household locations, so the (statistical) support is not spatially continuous \citep{Diggle2010}. 

We adopt an ordered approach: we first test for clustering (\S\ref{S:hypothesistesting}) and then, conditional on finding evidence against `no clustering' (nor inhibition), we estimate the \textit{clustering range} (\S\ref{S:parameterest}). We also provide the first precision estimate for the clustering range (see Graphical abstract). This approach is contrary to the current methods applied to the tau statistic and similar statistics \citep{Pollingtonreview}, which incorrectly combine graphical hypothesis testing for clustering and estimation of the clustering range (\S\ref{S:ourapp}). We hope these improved methods will encourage proper application of this burgeoning statistic.

\section{The tau statistic}
\label{S:thetaustatistic}
The tau statistic $\tau$ is a non-parametric global clustering statistic which takes a disease frequency measure (risk, odds or rate) within a certain annulus around an average case and compares it to the background measure (at any distance) \citep{Salje2012,Lessler2016,Pollingtonreview}. It measures the tendency of case pairs to spatially cluster while implicitly accounting for how related they are in terms of transmission using temporal information, making it a \textit{spatiotemporal} statistic. 

\subsection{Tau statistic (odds ratio estimator)}
\label{S:thetaustatisticodds}
We describe the most common tau estimator $\hat{\tau}_{\textnormal{odds}}$, which is based on the relative odds of disease \citep{Lessler2016}, rather than other forms of the statistic (including a new rate ratio estimator), which are described in a detailed review \citep{Pollingtonreview} from which this subsection draws heavily.

The distance form of the tau statistic $\tau_{\textnormal{odds}}$ is the ratio of i) the odds $\theta(d_l,d_m)$ of finding any case $j$ that is `related' to any other case $i$, within a half-closed annulus $[d_l, d_m)$, $(l,m\in \mathbb{Z}^{+}$, $l<m)$, around case $i$, to ii) the odds $\theta(0,\infty)$ of finding any case $j$ related to any case $i$ at any distance separation ($d_{ij}\geq 0$) for $n$ total cases (Equation \ref{eq:tauodds} \& Fig.\ \ref{fig:annulus}). 
\begin{equation}
\label{eq:tauodds}
\begin{split}
\hat{\tau}_{\textnormal{odds}}(d_l, d_m) &:= \frac{\hat{\theta}(d_l, d_m)}{\hat{\theta}(0,\infty)}\\
\textnormal{ where }&\hat{\theta}(d_l, d_m) = \frac{\sum_{i=1}^n\sum_{j=1, j\neq i}^n\mathds{1}(z_{ij} = 1, d_l\leq d_{ij}<d_m)}{\sum_{i=1}^n\sum_{j=1, j\neq i}^n\mathds{1}(z_{ij} = 0, d_l\leq d_{ij}<d_m)}
\end{split}
\end{equation}
The half-closed annulus is a correction to the original open interval (\citet{Lessler2016}: appendix 5); it was incorporated in December 2018 into the \texttt{IDSpatialStats} R package (which calculates the tau statistic) \citep{LesslerGiles}. The main computation of Equation \ref{eq:tauodds} is effectively a double sum over `relatedness' indicator functions $\mathds{1}(\cdot)$ for case pairs. $\hat{\tau}(d_l,d_m)$ is then evaluated over a \textit{distance band set} \underline{$\Delta$}. Sometimes an expanding disc is described by setting $d_l = 0$, relabelling $d_m=d$ to give $\hat{\tau}(d)$ instead. Although $\hat{\tau}$ is strictly evaluated for a given distance band $[d_l,d_m)$, when a $\tau$-distance graph is drawn a value of $\hat{\tau}(d)$ can be obtained through linearly interpolating between the distance band midpoints.

Tau values signify either the presence of spatiotemporal clustering ($\tau>1$), no clustering ($\tau = 1$) or inhibition ($\tau < 1$). The odds estimate $\hat{\theta}$ in Equation \ref{eq:tauodds} is the ratio of the number of related case pairs ($z_{ij}=1$) within $[d_1, d_2)$, versus the number of unrelated case pairs ($z_{ij}=0$) within $[d_l, d_m)$.

The relatedness of a case pair $z_{ij}$ is commonly determined using temporal information (e.g.\ difference in onset times of cases $i$ and $j$, i.e.\ $t_j-t_i$) \citep{Pollingtonreview}. The \textit{serial interval} is the period between the onset times of symptoms in the infector $t_i$ and their infectee $t_j$. Typically cases are defined as being temporally related when their onset times are within a single mean serial interval of each other.

\begin{center}
    * $^\tau$ *
\end{center}
In the following sections (\S\ref{S:methods}-\ref{S:resanddisc}) we provide a descriptive analysis of the data, before systematically testing several aspects of the tau statistic's implementation and their impact on the estimated clustering range.
\newpage

\section{Methods}
\label{S:methods}
\subsection{The dataset and baseline analysis}
\label{S:datasetandcomputmethods}
We analyse an infectious disease dataset of measles from case households in Hagelloch, Germany in 1861 \citep{Meyer2017,Neal2004,Oesterle1992,Pfeilsticker1863}. Computations were run in R using RStudio \citep{R,RStudio} with further detail in \ref{S:runcode} \& \ref{S:openaccess}. We have reproduced Lessler et al.'s (unpublished) analysis as a baseline result (Fig.\ \ref{fig:taurepro}). Using \textit{their} interpretation of Fig.\ \ref{fig:taurepro}, spatiotemporal clustering is reported up to 30m \citep{Lessler2016}. 

\subsection{Our approach to hypothesis testing and parameter estimation}
\label{S:ourapp}

An \textit{envelope} is loosely defined as a series of piecewise linear (syn. connected-line) functions in the Cartesian plane, with some bound applied above and below. \textit{Central}/\textit{null envelopes} describe the line function, i.e.\ whether it originates from simulations of a bootstrapped point estimate or time-permuted null distribution, respectively; whereas \textit{global envelope} or \textit{pointwise confidence interval} (syn.\ confidence band) refer to the way function lines are bounded. A global envelope is a confidence interval (CI) for a series of line functions but does not represent a single distance band of one tau point estimate $\hat{\tau}(d_l,d_m)$ (i.e.\ a pointwise CI), but rather the entire distance band set $\underline{\Delta}$. At say a 95\% significance level, in 95\% of outcomes of constructing a global envelope, the random envelope would contain the true value of $\tau(d_l,d_m), \forall\ [d_l,d_m)\in\underline{\Delta}$ \citep{Baddeley2015}.

Our graphical hypothesis test (\S\ref{S:hypothesistesting}) and parameter estimation (\S\ref{S:parameterest}) methods (see Graphical abstract) offer corrections to the implementations of the tau statistic or similar statistics used in many papers reviewed in \citet{Pollingtonreview} (i.e.\ \citet{Salje2012,Salje2016,Salje2016social,Salje2017,Salje2018,Grabowski2014,Bhoomiboonchoo2014,Levy2015,Lessler2016,Grantz2016,HoangQuoc2016,Succo2018,Rehman2018,Azman2018,Truelove2019}), which incorrectly used an envelope about the point estimate constructed from pointwise CIs or estimated the clustering endpoint $D$ as the distance at which the lower bound of the first pointwise percentile CI that is above $\tau =1$, touches $\tau =1$ (Fig.\ \ref{fig:incorrect}a) \citep{Pollingtonreview}. The former error amounts to multiple hypothesis testing and inflates type I errors (Fig.\ \ref{fig:incorrect}b).

\subsection{Graphical hypothesis test of no clustering}
\label{S:hypothesistesting}
We instead construct a \textit{global envelope} around the distribution of the null hypothesis ($H_0$: $\tau=1$, no spatiotemporal clustering) \nocite{Myllymakitalk}(Myllymäki, 2019b). This is generated by randomly permuting the time marks $t_i$ of the spatiotemporal data points $X_i$ = ($\textnormal{x-coordinate}_i$, $\textnormal{y-coordinate}_i$, $\textnormal{onset time}_i$) to scramble any spatiotemporal clustering present and simulate what $\hat{\tau}$ would be under $H_0$. We assess if a subset of distance bands $\underline{\delta}$ of $\underline{\Delta}$ exists (as contiguous or disjoint regions) where the tau point estimate $\hat{\tau}(d)$ is ever above/below the upper/lower bound, respectively, of this (global) null envelope. This null envelope is of extreme rank type (``defined as the minimum of pointwise ranks'') with 95\% significance level and extreme rank length $p\textnormal{-value}$ interval (note:\ a range, not a single $p\textnormal{-value}$) \nocite{Myllymaki2019}(Myllymäki et al., 2019a); as constructed by the \texttt{GET} R package \nocite{Myllymaki2019}(Myllymäki et al., 2019a) (see Graphical abstract). The test is two-tailed, which is necessary as only once the graph is plotted is the presence of clustering or inhibition known (alternative hypothesis $H_1:\tau\neq 1$). We compute 2,500 `time-mark permuted' tau simulations for an optimal test \citep{Myllymaki2017}.

\subsection{Parameter estimation of the clustering range}
\label{S:parameterest}
If hypothesis testing establishes the evidence against no spatiotemporal clustering within a subset of distance bands $\underline{\delta}$ (\S\ref{S:hypothesistesting}), it is then sensible to estimate the \textit{endpoint of spatiotemporal clustering} $\hat{D}$ for the clustering range $[d_1=0 \textnormal{ (assumed)}, d_{m}=\hat{D})$ where the point estimate intercepts $\tau = 1$, i.e.\ $\hat{D}:= \{d:\hat{\tau}(d)=1\}$. Due to discrete distance bands we linearly interpolate between the midpoint of distance band $[d_l,d_m)$ of the last $\hat{\tau}$ above one, and that of the next $\hat{\tau}$ below one $[d_{l+1},d_{m+1})$, to obtain $\hat{D}$.

To calculate the uncertainty of $\hat{D}$ we use bootstrapped tau estimates $\hat{\underline{\tau}}^*$. For each bootstrapped simulation (that represents a connected line of simulated tau estimates for increasing $d$ i.e.\ $\{\hat{\tau}^*(d_l,d_m):[d_l,d_m)\in\underline{\Delta}\}$), we record those that originate from above $\tau = 1$ and then intersect $\tau = 1$ at some greater distance $D$, i.e.\ those for which there exists $D$ satisfying $\hat{\tau}^*(D)=1$. We use $N=$ 2,500 samples which is more than sufficient for a typical bootstrap sample \citep{Efron1998}. We then take this horizontal set of values $\underline{D}$ and use it to obtain a CI to describe the uncertainty in $\hat{D}$ (see Graphical abstract). We now investigate \textit{spatial} bootstrap methods (\S\ref{S:bstrapmethod}), CI construction (\S\ref{S:CItype}) and distance band sets (\S\ref{S:dband}). 

\subsubsection{Spatial bootstrap sampling methods for $\hat{\tau}$}
\label{S:bstrapmethod}
To construct a central envelope for $\hat{\tau}$ we need to generate a non-parametric spatial bootstrap distribution of tau estimates, $\hat{\underline{\tau}}^*$. Through bootstrap theory, the sampling distribution $\hat{\underline{\tau}}^*$ may serve as a proxy for the actual distribution of $\hat{\tau}$ on the data; and further, the envelopes constructed from $\hat{\underline{\tau}}^*$ may approximate the envelope of $\hat{\tau}$ on the data \citep{Efron1979}. 
We compare three spatial bootstrap methods; all are non-parametric because they randomly resample the data without imposing a distribution \citep{Loh2008}.
\paragraph{Resampled-index spatial bootstrap (RISB)}
We start again with spatiotemporal data $\mathbf{X}=(X_i)_{i=1,\dots,n}$ where $X_i$ = ($\textnormal{x-coordinate}_i$, $\textnormal{y-coordinate}_i$, $\textnormal{onset time}_i$). Using the Uniform distribution we resample with replacement the data's indices $\underline{i} = (1, \ldots, n)$ $n$ times (equal to the number of cases), to produce a new empirical \textit{spatial bootstrap} sample of indices $\underline{i}^* = (i_k^*)_{k=1,\dots,n}$ and data $\mathbf{X}^* = \left(X_{\underline{i}^*}\right)$ ($\underline{i}$ and $\underline{i}^*$ have the same length, but $\underline{i}^*$ is bound to contain duplicated indices due to sampling with replacement). We compute the tau odds estimator on each bootstrap sample  $\mathbf{X}^*$ to get $N$ bootstrapped $\tau$ estimates $\hat{\underline{\tau}}^* = (\hat{\tau}^*_1, \ldots , \hat{\tau}^*_N)$; the same approach could be applied to other $\tau$ estimators. Loh critiques this ``naive'' sampling with replacement of the points $X_{\underline{i}}$ of a spatial dataset to produce a spatial bootstrap sample, because ``the spatial dependence structure has to be preserved as much as possible'' \citep{Loh2008} \ldots ``to reflect properties of the original process'' \citep{LohStein2004}. Lessler et al.\ and others used this method and additionally for any $p$, $q$ resampled indices ($p\neq q$), dropped $(i_p^*,j_q^*)$ pairs where they represented the same point ($i_p^*=j_q^*$) to avoid `self comparisons' \citep{Lessler2016}.

\paragraph{Modified marked point spatial bootstrap (MMPSB)}
Our method differs slightly to Loh \& Stein's (the second method here but discussed further in \S\ref{S:lohandsteinmpb}): rather than spatial bootstrapping the local $\tau$-functions (Equation \ref{eq:loh6}), we go deeper and compute the number of locally-related or locally-unrelated mark functions $m_i(k)$, according to their Boolean time-relatedness $k\in\{0, 1\}$. 

The number of time-related cases ($\#\textnormal{related}$) within a distance $[d_l, d_m)$ around a case $i^*$ chosen in the spatial bootstrap sample is:
\begin{equation}
\label{eq:loh1}
    \textnormal{\#related}(d_l,d_m,k=1,i^*) \equiv m_{i^*}(d_l,d_m,k=1) = \sum_{j\in\underline{j}, j\neq i^*}\mathds{1}(d_l\leq d_{i^*j} < d_m, z_{i^*j}=1)
\end{equation}
and then an average is taken over the $n$ cases in the spatial bootstrap sample of indices $\underline{i}^*$:
\begin{equation}
\label{eq:loh2}
    \overline{\textnormal{\#related}^*(d_l,d_m)} \equiv m^*(k=1) = \frac{1}{n}\sum_{i^*\in\underline{i}^*}\sum_{j\in\underline{j},j\neq i^*}\mathds{1}(d_l\leq d_{i^*j} < d_m, z_{i^*j}=1),
\end{equation}
and similar steps for time-unrelated cases yield:
\begin{equation}
\label{eq:loh3}
    \overline{\textnormal{\#unrelated}^*(d_l,d_m)} \equiv m^*(k=0) = \frac{1}{n}\sum_{i^*\in\underline{i}^*}\sum_{j\in\underline{j},j\neq i^*}\mathds{1}(d_l\leq d_{i^*j} < d_m, z_{i^*j}=0),
\end{equation}
and finally the odds and odds ratio estimator can be calculated as before:
\begin{equation}
\label{eq:loh4}
    \theta^*(d_l,d_m) = \frac{\overline{\textnormal{\#related}^*(d_l,d_m)}}{\overline{\textnormal{\#unrelated}^*(d_l,d_m)}} = \frac{\sum_{i^*\in\underline{i}^*}\sum_{j\in\underline{j},j\neq i^*}\mathds{1}(d_l\leq d_{i^*j} < d_m, z_{i^*j}=1)}{\sum_{i^*\in\underline{i}^*}\sum_{j\in\underline{j},j\neq i^*}\mathds{1}(d_l\leq d_{i^*j} < d_m, z_{i^*j}=0)}
\end{equation}
\begin{equation}
\label{eq:loh5}
    \tau_{\textnormal{MMPSB}}^*(d_l,d_m) = \frac{\theta^*(d_l,d_m)}{\theta^*(0,\infty)}
\end{equation}

\subsubsection{Confidence interval (CI) construction}
\label{S:CItype}
Applying a percentile CI to the sample bootstrap distribution $\underline{D}$ (previously defined in \S\ref{S:parameterest}) assumes it is symmetric which is not the case, especially at short distances (Fig.\ \ref{fig:bootstraphists}) \citep{Carpenter2000}. 

Bias-corrected and accelerated (BCa) CIs can cope with asymmetrical distributions better than percentile CIs. For non-parametric problems \citet{Carpenter2000} consistently found Efron's BCa method best due to its low theoretical coverage errors for approximating the exact CI. BCa had ``\textit{second-order correct coverage}'' errors under some assumptions, while a percentile CI was first-order correct at best \citep{Efron1987}. The BCa algorithm transforms a distribution of bootstrap calculations by normalisation to stabilise its variance so that a CI can be constructed, then back-transforms it \citep{Efron1987}. We calculated it using the \texttt{coxed} R package \citep{coxed}.

\subsubsection{Distance band sets}
\label{S:dband}
The tau statistic is non-unique as it depends on the distance band set chosen \citep{Pollingtonreview}, so the potential variation in $\tau$ estimates from this choice is of interest. From analysing cases' pairwise distances we propose a reasonable non-overlapping distance band set, i.e.\ $\underline{\Delta}=$ \big\{[0,7), [7,15), [15,20), [20,25), [25,30), \ldots, [195,200m)\big\} as a comparison to Lessler et al.'s overlapping set \big\{[0,10), [0,12), [0,14), \ldots, [0,50), [2,52), [4,54), \ldots, [74,124m)\big\}, and test these using $N=$ 2,500 samples under the MMPSB method.
\newpage

\section{Results \& discussion}
\label{S:resanddisc}

\subsection{Dataset description}
\label{S:datasetdesc}
The epidemic over a small $\sim$280m x 240m area lasted nearly three months and five distinct generations can be discerned from the epidemic curve (Fig.\ \ref{fig:re}). Out of the 197 under-14 year olds, 185 became infected, along with three teenagers, leaving 377 remaining teenagers and adults uninfected \citep{Neal2004}. Figure \ref{fig:stplot} indicates a weak signal of direct transmission between cases, as cases with onsets close together in time (shown by similar colours) tended to be nearby to each other.

\subsection{Graphical hypothesis tests: global envelopes vs pointwise CIs}
\label{S:ResDiscGET}
There is moderately strong evidence against the hypothesis of no spatiotemporal clustering ($p\textnormal{-value}$ $\in [0, 0\text{\textperiodcentered}014]$) based on constructing the global envelope around $\tau=1$ under the null hypothesis (Fig.\ \ref{fig:get}), and thus we conclude that the data $\mathbf{X}$ is inconsistent with the null model ($H_0: \tau=1$). So we turn to the alternative hypothesis, that there is clustering and/or inhibition. Fig.\ \ref{fig:get} suggests there is clustering at short distances and inhibition at long distances. Unfortunately it is not possible to compare our results with those of previous papers (see \S\ref{S:ourapp}), since they used an incorrect pointwise CI approach to assess clustering, for which a $p\textnormal{-value}$ is not available.

\subsection{Impact on the estimated clustering endpoint}
\label{S:nbootstrapsres}
The estimated clustering endpoint is $\hat{D}=$ 36\text{\textperiodcentered}0m with a 95\% percentile CI of (14\text{\textperiodcentered}5, 58\text{\textperiodcentered}0m) over 100 bootstrapped simulations using RISB sampling (Fig.\ \ref{fig:nbootstraps}), or (14\text{\textperiodcentered}6, 58\text{\textperiodcentered}5m) over 2,500 simulations (using 100\% of simulations, see \ref{S:appinvalidate}); more bootstrapped simulations do not appear to affect the precision. 

The point estimate $\hat{D}=$36\text{\textperiodcentered}0m is only 20\% higher than the baseline clustering range (30m). Previous estimates derived via the improper method of finding the distance at which the lower bound of the central envelope (around $\hat{\tau}$) touches $\tau = 1$ underestimated this range. The plateauing shape of $\hat{\tau}(d)$ before it reaches $\tau = 1$ contributes to the increased imprecision in the estimate of $\hat{D}$. This highlights the utility of a human assessing the graph rather than rigidly using a $\tau = 1$ threshold, as it is likely that disease control over say a 60m radius around an average case would see the biggest gains over its first 15 metres with diminishing returns at wider radii (Fig.\ \ref{fig:nbootstrapsv2}). 

\subsection{Spatial bootstrap sampling: modified marked point vs resampled-index}
Using the modified marked point spatial bootstrap (MMPSB) (\S\ref{S:bstrapmethod}) yields a narrower envelope than the resampled-index spatial bootstrap (RISB), leading to a 95\% BCa CI for $\hat{D}$ of (14\text{\textperiodcentered}9, 46\text{\textperiodcentered}6m) (Fig.\ \ref{fig:nbootstrapsv2}); both CIs used 100\% of simulations. 

If the tau point estimate had been shallower near the $\tau=1$ intercept then the range of spatiotemporal clustering would be far larger and the benefit of MMPSB more apparent. Given the reasons why this method is better (\S\ref{S:bstrapmethod}), we believe the RISB will underestimate this range. 

The MMPSB outperforms the RISB because the latter loses more pair information from resampling indices and avoiding self-comparisons. This was checked empirically for the measles data: the tau point estimate was computed on 188 x 187 = 35,156 pairs. On average from 1,000 simulations, the RISB sampled from 119 unique people, leading to 119 x 118 = 14,042 unique pairs evaluated or $\sim 39\text{\textperiodcentered}9$\% of the original pairs. Of course many additional duplicate pairs are used in the RISB but we are only interested in unique pair information that is retained. The MMPSB only has 119 unique mark functions, but each of them is compared with the other 187 cases, leading to 63\text{\textperiodcentered}3\% of pairs being retained.

\subsection{Confidence interval: BCa vs percentile}
\label{S:bcaresults}
Histograms of the asymmetric distribution of $\underline{D} = \{D_i: \hat{\tau}^*_i(D_i)=1, i=1,\dots,N\}$ by number of bootstrapped samples indicate for both $N=$ 100 or 2,500 samples that a percentile CI gives a less precise estimate; both CIs used 100\% of simulations (Fig.\ \ref{fig:bootstraphists}). The BCa method provides slightly narrower CIs than the original percentile CIs (Fig.\ \ref{fig:bootstraphists}). The RISB appears to introduce positive skew (mean $>$ median) in $\underline{D}$ whereas MMPSB with sufficient samples ($N=2500$) introduces a slight negative skew. MMPSB reduces the bias, $\underline{\bar{D}}-\hat{D}$, between mean/median estimates of $\underline{D}$ and the point estimate $\hat{D}$ from $\sim$10m to $\sim$5m, or $\sim$17\% of $\hat{D}$.

\subsection{Distance bands}
Overlapping distance band sets appear to produce $\hat{D}$ estimates with more variance (95\% BCa CI (14\text{\textperiodcentered}9, 46\text{\textperiodcentered}6m)) than non-overlapping sets (CI (15\text{\textperiodcentered}4, 26\text{\textperiodcentered}1m)) (Fig.\ \ref{fig:distband}), but a clearer and smoother trend in tau with increasing distance (both CIs used 100\% of simulations). The non-overlapping $\underline{\Delta}$ also struggles to contain $\hat{D}$ (Fig.\ \ref{fig:distband}) because the simulations are more erratic about $\tau = 1$, the distribution of $\underline{D}$ is strongly bi-modal, which even the BCa technique cannot account for. The increased volatility of $\hat{\tau}$ also results in multiple intercepts with $\tau = 1$, but for usability we prefer a single range of clustering, given in this case by the overlapping $\underline{\Delta}$.

\begin{center}
    * $^\tau$ *
\end{center}

The 20\% increase in the radial parameter $\hat{D}$ (\S\ref{S:nbootstrapsres}) from using the corrected parameter estimation algorithm (\S\ref{S:parameterest}) may not seem important for public health interventions, but their time and cost is more closely proportional to area, and the areal increase is 44\% (since  $\pi(1\text{\textperiodcentered}20\hat{D})^2/\pi \hat{D}^2=1\text{\textperiodcentered}44$, assuming $d_1=0$).
\newpage

\section{Conclusion and recommendations for improved use}
\label{S:concl}

We have shown that the way clustering ranges are currently calculated using the tau statistic can lead to biased estimates. Using a modified marked point spatial bootstrap and BCa CIs to calculate the clustering range for the Hagelloch measles dataset resulted in bias reductions equivalent to increasing the clustering area of elevated odds by $44\%$. These improvements will appear in future versions of the \texttt{IDSpatialStats} package. Our results (\S\ref{S:resanddisc}) support the following recommendations: 
\begin{itemize}
\item the modified marked point spatial bootstrap should be used to simulate $\hat{\tau}$ instead of the resampled-index method that could lead to underestimation of the clustering range.
\item BCa, rather than percentile, CIs should be used as they give better coverage when the bootstrap distribution of tau simulations $\hat{\underline{\tau}}^*$ is non-symmetric. 
\end{itemize}

\paragraph{Tau statistic limitations} 
The distance band set choice $[d_l,d_m)\in\underline{\Delta}$ clearly affects the smoothness of the point estimate $\hat{D}$, and its precision. A better understanding of how to choose distance bands for a given purpose is now needed. It is also unknown how the time-relatedness interval choice $[T_1,T_2]$ (where $z_{ij} = \mathds{1}\big((t_j-t_i)\in [T_1,T_2]\big)$) biases the tau statistic through inclusion of extraneous co-primary or secondary cases. It is unclear how second-order correlation functions like the tau statistic and Ripley's $K$ function \citep{Gabriel2009}, originally founded in spatiotemporal point processes with continuous support in $\mathbb{R}^2$, behave for this data. Finally the number of bootstrap samples required for graphical hypothesis testing and estimation purposes is unknown; we believe that related research by \citet{Davidson2000} could inform a heuristic algorithm.
\begin{center}
    * $^\tau$ *
\end{center}
We encourage the adoption of the statistical protocol described (see Graphical abstract) to properly test for clustering, and, if appropriate, estimate its range. Control programmes are being informed by the tau statistic and applying these bias-reduction methods will improve its accuracy and future health policy decisions. In addition to modellers or epidemiologists working on real-time outbreaks or post-study analysis, we hope statisticians are inspired to apply this statistic to spatiotemporal branching processes in new fields.
\newpage

\section{Acknowledgements}
\label{S:ack}
TMP would like to thank:
\begin{itemize}
    \item Henrik Salje and Justin Lessler for sharing unpublished analysis code for our reproduction \citep{Lessler2016}, and Shaun Truelove, John Giles and Henrik Salje for useful discussions at a poster presentation \citep{Pollingtonposter}.
    \item Peter Diggle (PJD) for highlighting an earlier spurious result, a mistake in conflating parameter estimation with hypothesis testing, and for reviewing our second draft.
    \item Mari Myllymäki for \texttt{GET} R package support.
    \item The Editors and Reviewers of \textit{Spatial Statistics} who made pertinent suggestions.
\end{itemize}

TMP, LACC \& TDH gratefully acknowledge funding of the NTD Modelling Consortium by the Bill \& Melinda Gates Foundation (BMGF) (grant number OPP1184344) and LACC acknowledges funding of the SPEAK India consortium by BMGF (grant number OPP1183986). Views, opinions, assumptions or any other information set out in this article should not be attributed to BMGF or any person connected with them. TMP's PhD is supported by the Engineering \& Physical Sciences Research Council, Medical Research Council and University of Warwick (grant number EP/L015374/1). TMP thanks Big Data Institute for hosting him during this work. All funders had no role in the study design, collection, analysis, interpretation of data, writing of the report, or decision to submit the manuscript for publication.

\section{Competing interests}
All authors declare no competing interests.

\section{Contributions: CRediT statement}
\noindent \textbf{TMP:} Conceptualisation, Methodology, Software, Validation, Formal analysis, Investigation, Data curation, Writing - original draft \& editing, Visualisation \textbf{MJT:} Conceptualisation, Writing - review \& editing, Supervision \textbf{PJD:} Methodology, Validation (see \S\ref{S:ack}), Writing - review \& editing \textbf{TDH:} Conceptualisation, Writing - review \& editing, Supervision, Funding acquisition \textbf{LACC:} Conceptualisation, Software, Validation, Data curation, Writing - review \& editing, Supervision.

\section{Open access}
\label{S:openaccess}
The analysis code in \texttt{R Markdown} is available from \url{https://github.com/t-pollington/developments_tau_statistic} under a GNU General Public License v3·0 licence. This article is licensed under the Creative Commons Attribution-NonCommercial-NoDerivatives Works 4·0 International Licence (CC BY-NC-ND 4·0). Anyone can copy and distribute this article unchanged and unedited but only for non-commercial purposes, provided the user gives credit by providing this article's DOI and a link to the licence (\url{creativecommons.org/licences/by-nc-nd/4.0}). The use of this material by others does not imply the authors' endorsement.
\newpage

\bibliographystyle{apa}
\bibliography{sample.bib}
\newpage

\section{Figures}
\begin{figure}[!h]
\centering\includegraphics[width=1.0\linewidth]{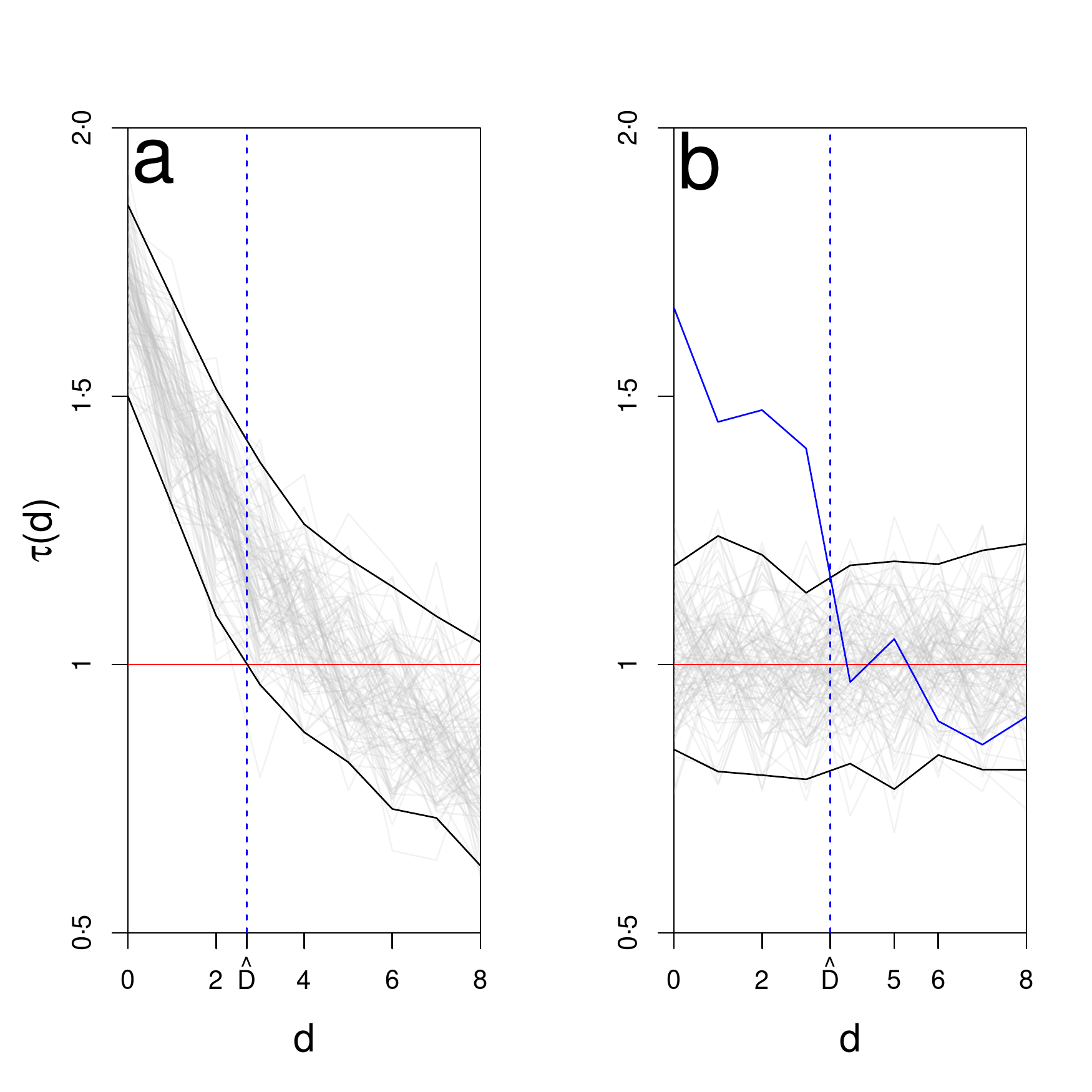}
\caption{The naïve methods employed by several authors (\S\ref{S:ourapp}) who choose one envelope type as `central' (a) or `null' (b), then simultaneously test the hypothesis of clustering and estimate the range of clustering parameter $\hat{D}$ \citep{Pollingtonreview}. The single red line $\tau = 1$ represents no spatiotemporal clustering. Grey lines indicate a) negative exponential lines with Normal noise to characterise a series of spatial bootstrap estimates $\hat{\tau}^*$ of a typical tau function, or b) a line at $\tau=1$ with Normal noise to represent simulations of $\tau = 1$ for null envelope construction; black lines mark out the envelope bounds. The solid blue line characterises an empirical tau point estimate $\hat{\tau}(d)$. Instead, we split the method into separate hypothesis testing and parameter estimation steps in $\S\ref{S:hypothesistesting}$ \& $\S\ref{S:parameterest}$, respectively.}
\label{fig:incorrect}
\end{figure}

\begin{figure}[!h]
\centering\includegraphics[width=1.0\linewidth]{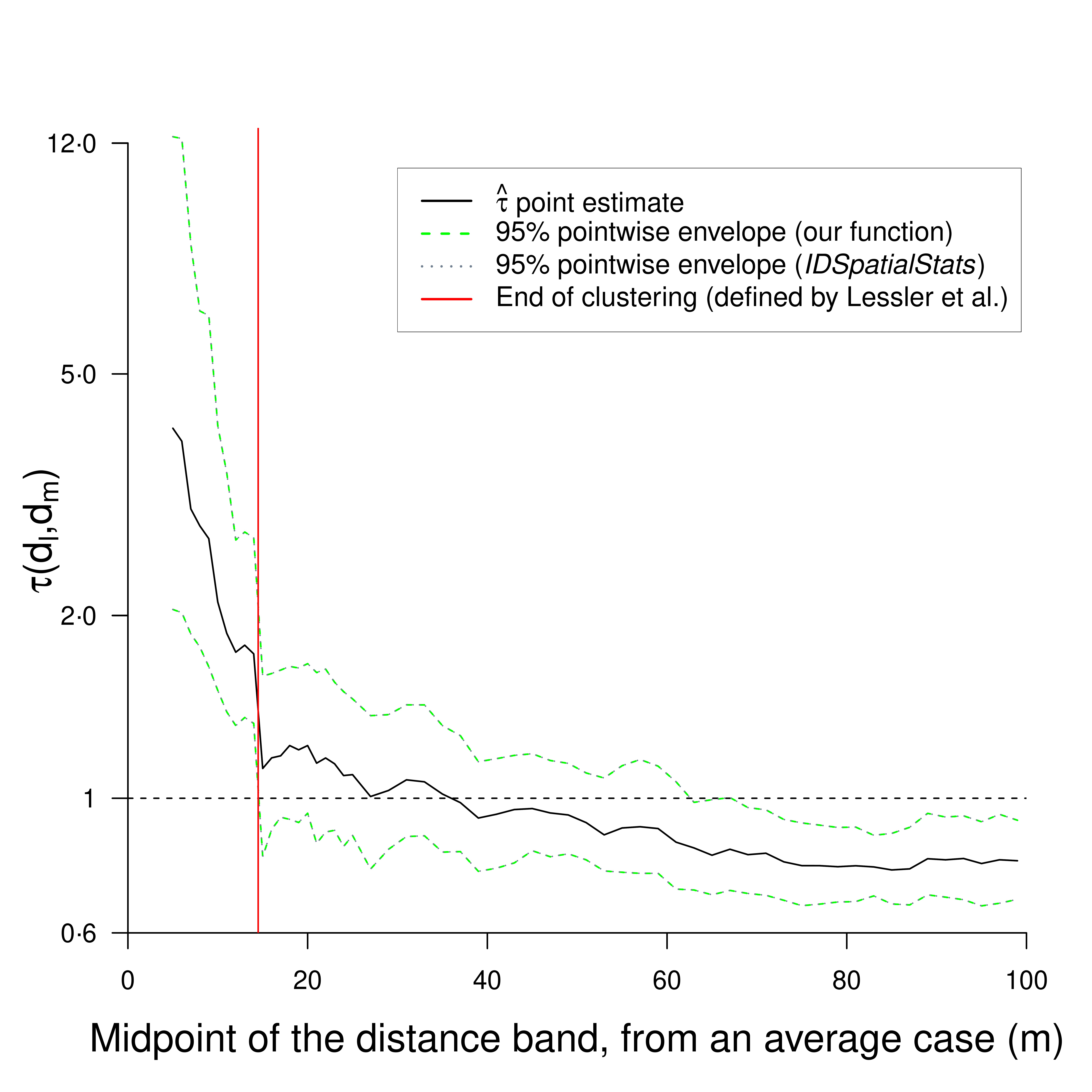}
\caption{Baseline result: a reproduction of a previous analysis \cite[Fig. 4C]{Lessler2016}. Note that the end of the clustering range reported by Lessler et al.\ is where the lower bound of the envelope intersects $\tau = 1$ (14\text{\textperiodcentered}5m or 15m rounded) (we do \emph{not} endorse this convention however). Regardless, as the horizontal axis is the midpoint of the distance band (i.e.\ $(d_1+d_2)/2$), [0, 30)m is the actual clustering range that would be interpreted using their convention, as confirmed by Lessler (Lessler, personal comm.). The near perfect superimposition of their envelope and ours validates our implementation of tau functions from their \texttt{IDSpatialStats} R package.}
\label{fig:taurepro}
\end{figure}

\begin{figure}[!h]
\centering\includegraphics[width=1.0\linewidth]{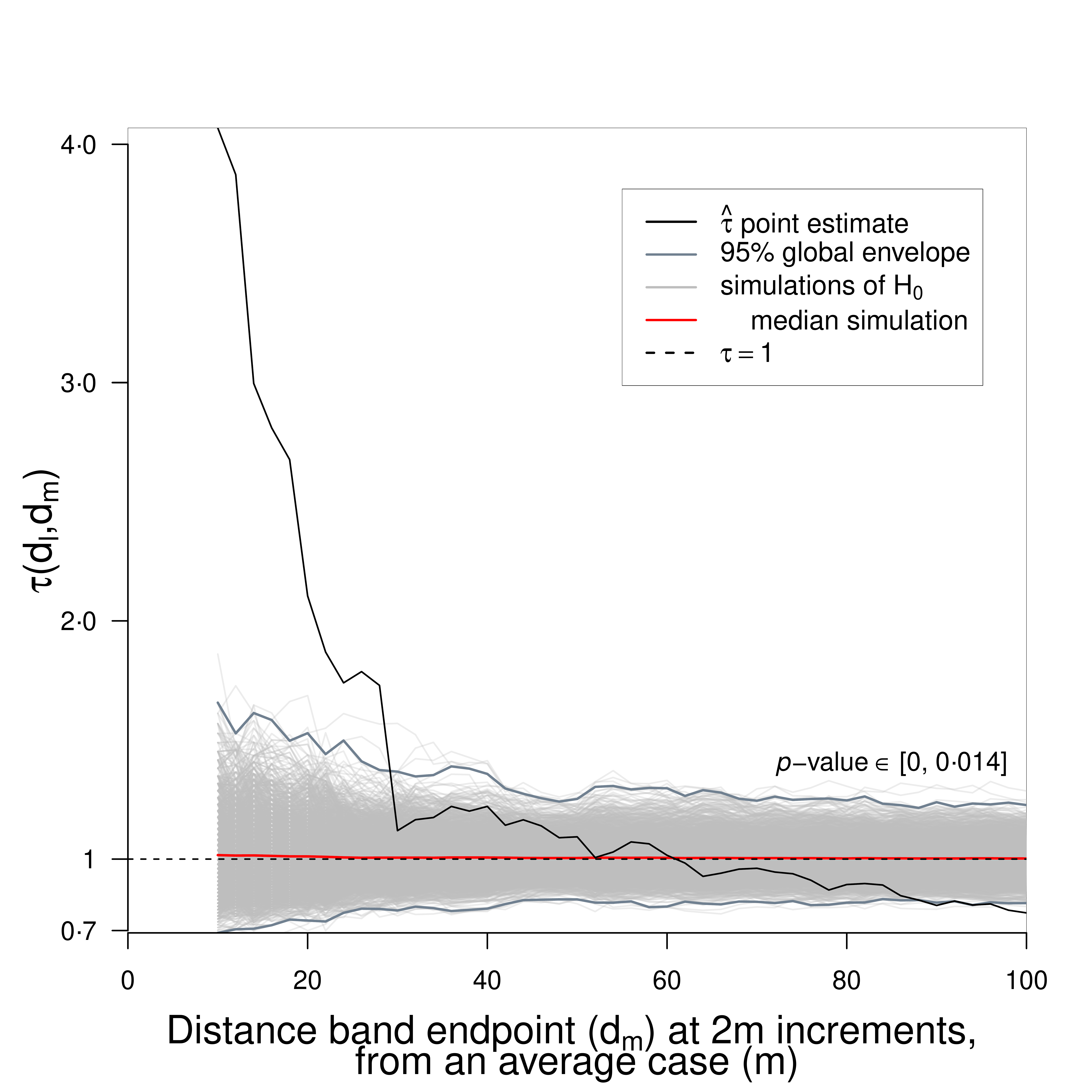}
\caption{Global envelope test, `extreme rank' type, two-sided at 95\% significance level using 2,500 simulations of the null hypothesis ($H_0$: no spatiotemporal clustering, i.e.\ $\tau= 1$). Note there is a region where $\hat{\tau}$ just exits the global envelope lower bound (suggesting inhibition at long distances) as well as the obvious departure above the upper bound (suggesting clustering at close distances). We are confident that we are simulating $H_0$ because the median simulation stays close to $\tau = 1$ throughout. Distance band set $:=\big\{[0, 10), [0, 12), [0,14), \ldots, [0, 50), [2, 52), [4, 54), \ldots, [74, 124)\textnormal{m}\big\}$.}
\label{fig:get}
\end{figure}

\begin{figure}[!h]
\centering\includegraphics[width=1.0\linewidth]{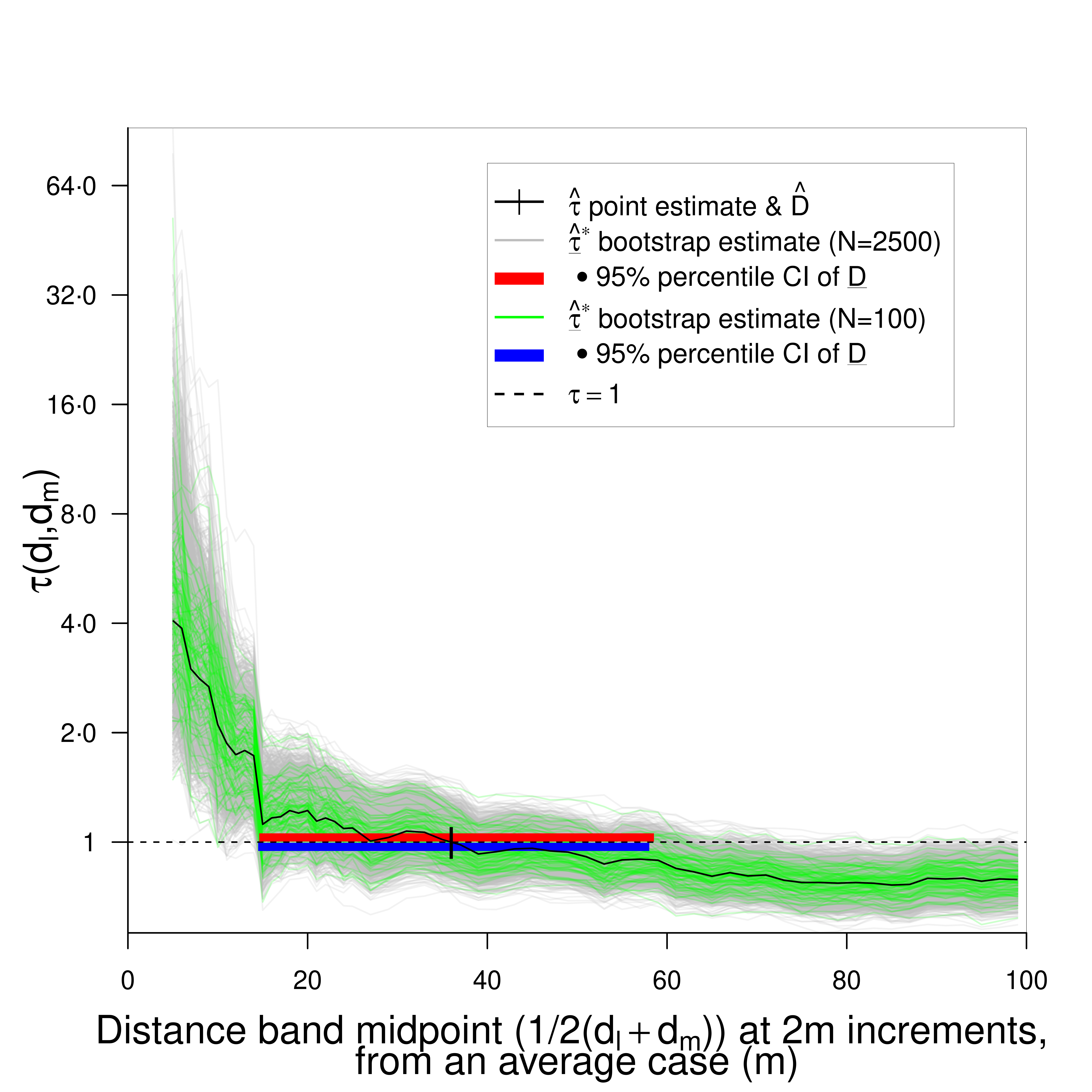}
\caption{Effect of number of samples on $\hat{D}$ precision, when using RISB sampling. Both CIs used 100\% of simulations. $\hat{D}=36\text{\textperiodcentered}0$m; $N=100$: 95\% BCa CI (14\text{\textperiodcentered}5, 58\text{\textperiodcentered}0m); $N=2500$: CI (14\text{\textperiodcentered}6, 58\text{\textperiodcentered}5m). Distance band set as Fig.\ \ref{fig:get}.}
\label{fig:nbootstraps}
\end{figure}

\begin{figure}[!h]
\centering\includegraphics[width=1.0\linewidth]{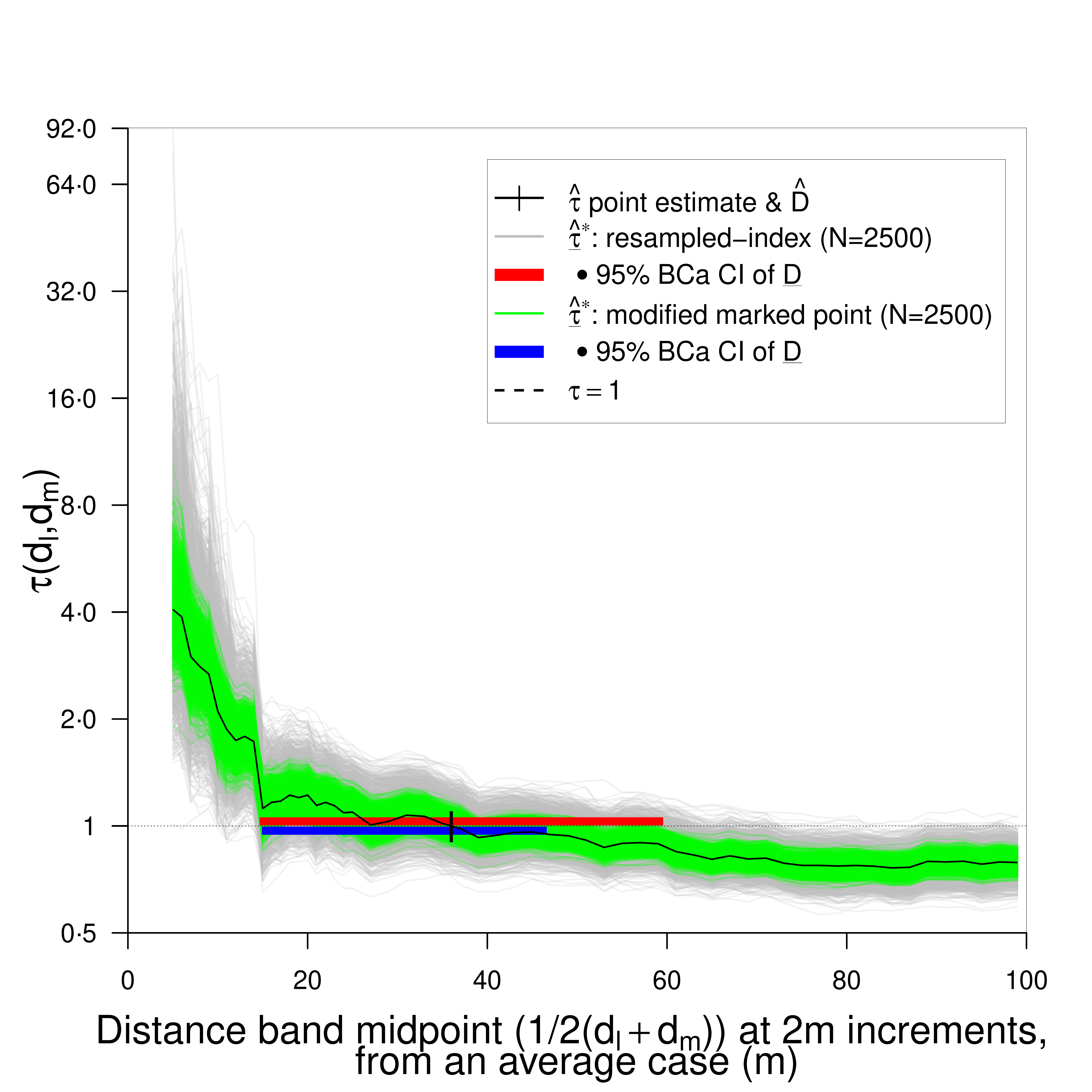}
\caption{Effect of spatial bootstrap sampling method on $\hat{D}$ precision. RISB 95\% BCa CI (14\text{\textperiodcentered}7, 60\text{\textperiodcentered}0m); MMPSB CI (14\text{\textperiodcentered}9, 46\text{\textperiodcentered}6m); both CIs used 100\% of simulations. Distance band set as Fig.\ \ref{fig:get}, $N=$2500. }
\label{fig:nbootstrapsv2}
\end{figure}

\begin{figure}[!h]
\centering\includegraphics[width=0.85\linewidth]{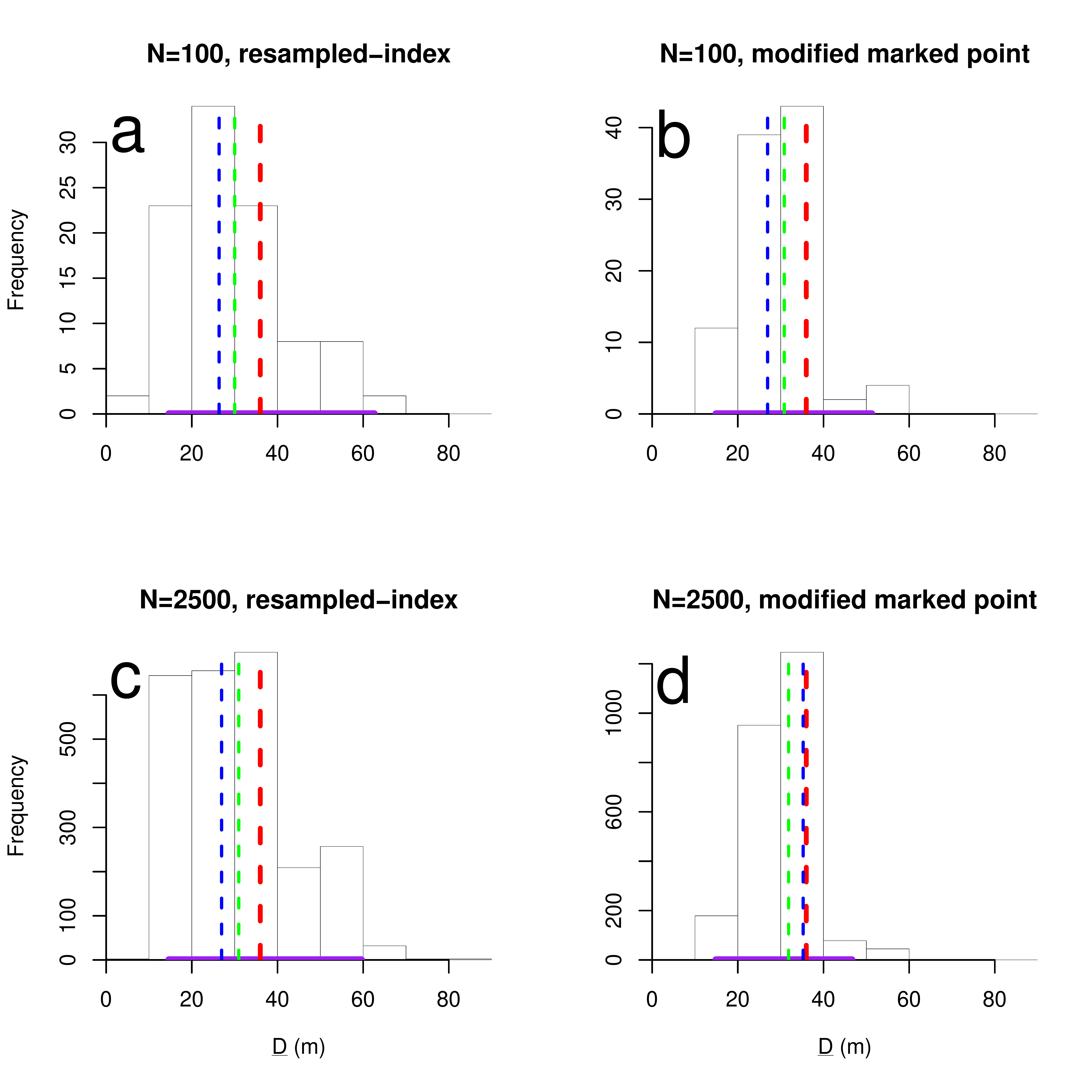}
\caption{Distribution of $\underline{D}$, the set of samples from the sampling distribution of values of $\hat{D}$, i.e\ $\underline{D}=\{\hat{D}_i:\hat{\tau}^*_i(\hat{D}_i)=1,i=1,\dots,N\}$ (illustrated in Graphical abstract), by number of bootstrapped samples N=100 (top row) or N=2500 (bottom) and by spatial bootstrap sampling method RISB (left column) or MMPSB (right). Vertical dotted lines indicate the $\hat{\tau}$ point estimate (red), mean (green) and median (blue) of the bootstrapped tau estimates. For the RISB both have positive skew as the mean estimate is greater than the median estimate, whereas for the MMPSB both have a negative skew. All spatial bootstrap estimations have a negative bias with respect to mean or median summary measures versus the point estimate, of approximately $\sim$10m for the RISB and approximately $\sim$5m for the MMPSB. The data points used to construct the BCa CIs (purple line on horizontal axis) from the $\hat{D}$ estimates in (a) are copied from Fig.\ \ref{fig:nbootstraps} (N=100 simulations) while those for (c) \& (d) are from Fig.\ \ref{fig:nbootstrapsv2}, while (b) has been freshly calculated. All four CIs used 100\% of simulations. Distance band set as Fig.\ \ref{fig:get}.} 
\label{fig:bootstraphists}
\end{figure}

\begin{figure}[!h]
\centering\includegraphics[width=1.0\linewidth]{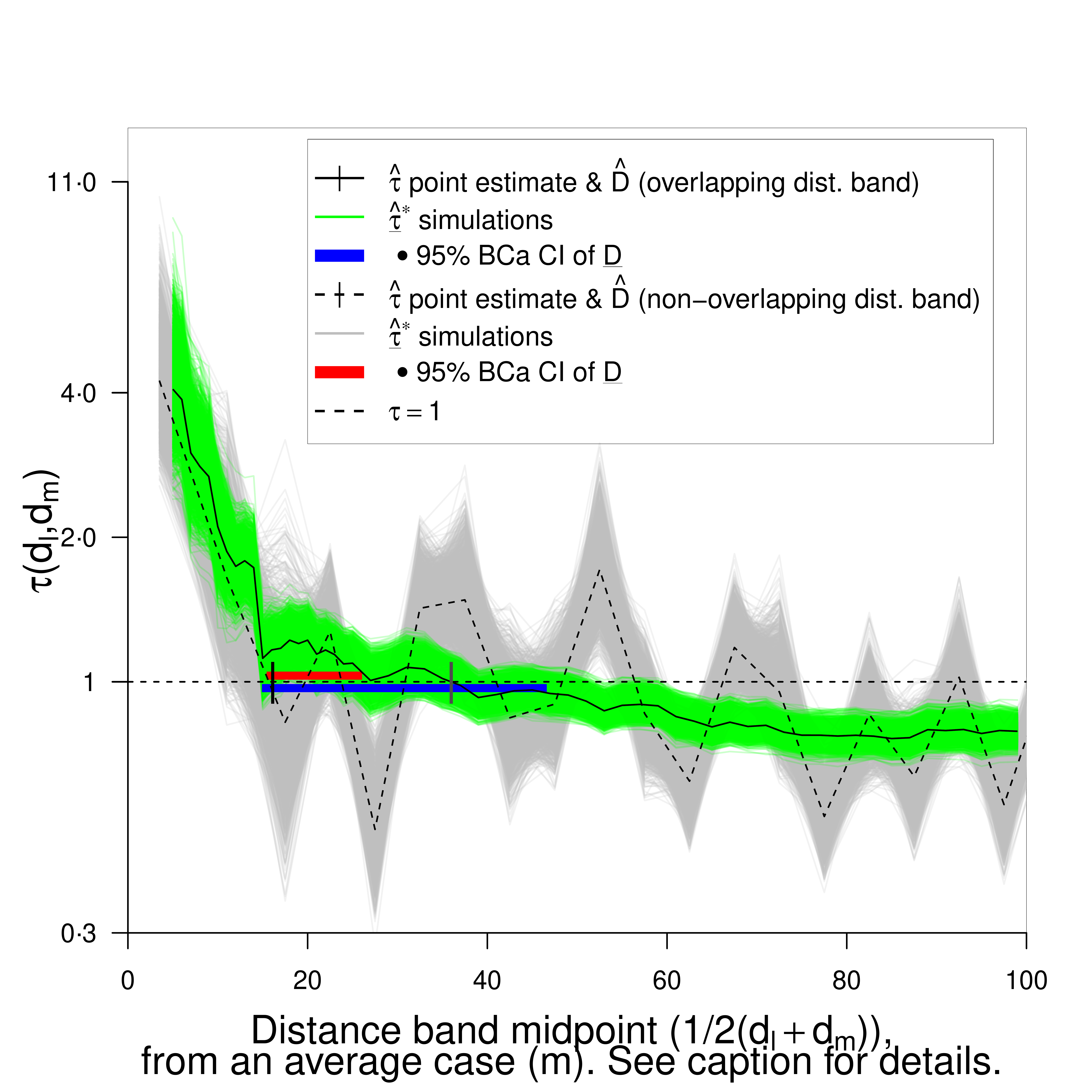}
\caption{Effect of distance band set on $\hat{D}$ precision using MMPSB sampling. Overlapping set (Lessler et al.) $:=\big\{[0, 10), [0, 12), [0,14), \ldots, [0, 50), [2, 52), [4, 54), \ldots, [74, 124)\textnormal{m}\big\}$ and non-overlapping $:=\big\{[0, 7), [7, 15), [15,20), [20,25), [25,30), \ldots, [195, 200)\textnormal{m}\big\}$. Non-overlapping sets yield a more erratic point estimate $\hat{\tau}$ yet tighter 95\% BCa CI (15\text{\textperiodcentered}4, 26\text{\textperiodcentered}1m) versus (14\text{\textperiodcentered}9, 46\text{\textperiodcentered}6m) however on further investigation the distribution of $\underline{D}$ is heavily bimodal; both CIs used 100\% of simulations.}
\label{fig:distband}
\end{figure}

The following figures cover illustration (Fig.\ \ref{fig:annulus}), previous methods and the baseline analysis (Figs.\ \ref{fig:incorrect} \& \ref{fig:taurepro}), perform a global envelope test (Fig.\ \ref{fig:get}), or investigate effects on the parameter point estimate $\hat{D}$ and distribution $\underline{D}$ (unless stated the distance band set is `overlapping', see Fig.\ \ref{fig:get} caption): 
\begin{itemize}
\item number of samples ($N=$100 or $N=$2500) on RISB sampling using percentile CIs (Fig.\ \ref{fig:nbootstraps})
\item (RISB vs.\ MMPSB) or (MMPSB vs.\ MPSB) sampling, using $N=$2500 and BCa CIs (Figs.\ \ref{fig:nbootstrapsv2} \& \ref{fig:loh}, respectively)
\item RISB vs.\ MMPSB sampling and ($N=$100 or $N=$2500) (Fig.\ \ref{fig:bootstraphists})
\item overlapping vs.\ non-overlapping distance band sets, using $N=$2500, MMPSB sampling and BCa CIs (Fig.\ \ref{fig:distband})
\end{itemize}

\begin{figure}[!h]
\centering\includegraphics[width=0.44\linewidth]{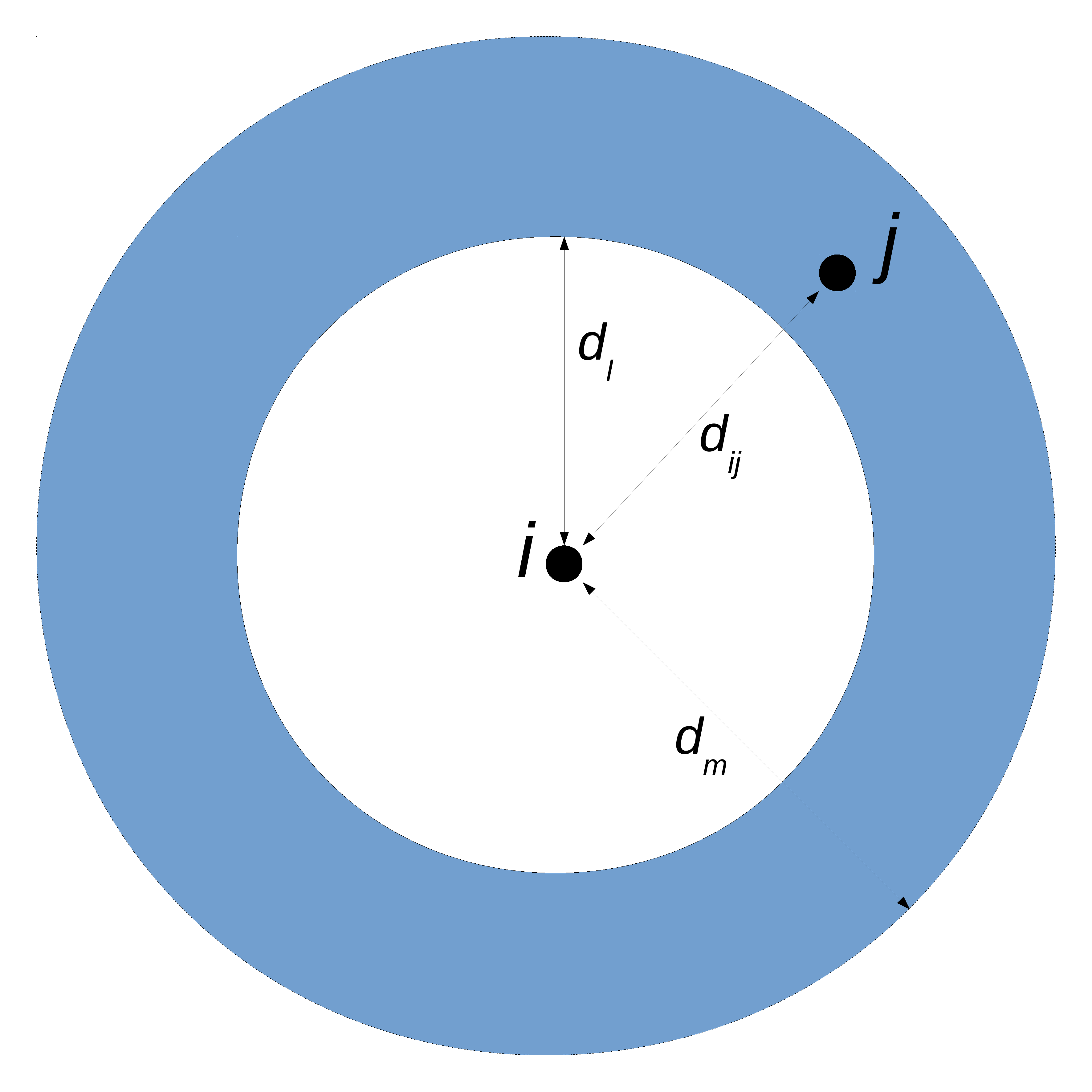}
\caption{A single distance band half-closed annulus of radii $[d_l, d_m)$ around an average case $i$ with another case $j$ in it, separated by distance $d_{ij}$.}
\label{fig:annulus}
\end{figure}

\newpage
\FloatBarrier

\appendix
\section{Extended notes on Methods section (\S\ref{S:methods})}
\subsection{Computation methods}
\label{S:runcode}
The \texttt{spatstat} library \citep{spatstat} was used for useful spatial functions, \texttt{purrr} for resampling \citep{purrr}, \texttt{fields} for image plots \citep{fields} and \texttt{latex2exp} \& \texttt{scales} for graph notation \citep{latex2exp,scales} and the code of `January' \nocite{January2017}(2017) for figure labelling. The \texttt{IDSpatialStats::get.tau()} and \texttt{get.tau.bootstrap()} functions were optimised by re-implementing them in C, which sped up $\tau_{\textnormal{odds}}$ calculations by $\sim$29 times \citep{Pollington2019}. For consistency we used Lessler et al.'s overlapping distance band set throughout, i.e.\ $\underline{\Delta} = $\big\{[0, 10), [0, 12), [0,14), \ldots, [0, 50), [2, 52), [4, 54), \ldots, [74, 124)\big\}.

\subsection{Invalidation of the confidence interval for the endpoint of spatiotemporal clustering}
\label{S:appinvalidate}

The confidence interval (CI) for the endpoint of spatiotemporal clustering $\hat{D}$ is easily invalidated if not all $\hat{\tau}^*$ simulations intersect $\tau = 1$ within the distance band set $\underline{\Delta}$. Caution is needed as the simulations $\underline{D}$ on which the uncertainty in $\hat{D}$ is calculated, are not a random sample of the population of simulations $\underline{\hat{\tau}}^*$, which is an important prerequisite for CI construction, as we selectively choose those that cross $\tau = 1$ from above and ignore those that start at or below $\tau=1$, or above it but never reach $\tau = 1$. Computing CIs at a 95\% confidence level on any random sample with a small 5\% dropout can substantially decrease the effective confidence level \citep{Gorard2014}. This selection bias is also $\underline{\Delta}$-dependent since if we choose a large enough $\underline{\Delta}$, we may find that simulations that start above $\tau=1$ eventually do cross $\tau = 1$ and then contribute to the CI. 
Although we cannot account for this bias, we do report the proportion of simulations used to construct the CIs and extend the distance range as computation time permits, to limit this bias.

\subsection{Estimating the startpoint of spatiotemporal inhibition}
If inhibition was present at greater distances we ignored estimating its range as it was not of interest. However, if the reader wishes, it can be estimated using a similar algorithm as for estimating clustering at shorter distances, in which one instead captures simulation lines that \emph{exit} the global envelope lower bound into $\tau < 1$ for increasing $d$. 

\subsection{Loh \& Stein marked point spatial bootstrap (MPSB) applied to the tau odds ratio estimator, not recommended}
\label{S:lohandsteinmpb}
The Loh \& Stein marked point spatial bootstrap (MPSB) is a fast, non-parametric method to obtain a bootstrap distribution of a second-order correlation function \citep{LohStein2004}. For a clustered process simulated by a Mat\'{e}rn process, the CIs constructed using it had a higher empirical coverage than other methods, and computed faster \citep{LohStein2004}.

For the RISB (\S\ref{S:bstrapmethod}) each bootstrap estimate $\hat{\tau}^*$ is computed from resampled (and smaller) spatiotemporal data $\mathbf{X}^*$ containing duplicated points from duplicate indices in $\underline{i}^*$, but the MPSB instead takes a spatial bootstrap sample of the locally-evaluated $\tau$-functions $\underline{\tau}_i$ (Equation \ref{eq:lohandstein}) corresponding to each $i^*\in \underline{i}^*$ across all points $\underline{j}, j\neq i^*$, so each local $\tau_i$ covers all points in $\mathbf{X}^*$ unlike the RISB:
\begin{equation}
    \label{eq:lohandstein}
    \begin{split}
    \hat{\tau}_{i}(d_l, d_m) &:= \frac{\hat{\theta_i}(d_l, d_m)}{\hat{\theta_i}(0,\infty)}\\
    \textnormal{ where }&\hat{\theta_i}(d_l, d_m) = \frac{\sum_{j=1, j\neq i}^n\mathds{1}(z_{ij} = 1, d_l\leq d_{ij}<d_m)}{\sum_{j=1, j\neq i}^n\mathds{1}(z_{ij} = 0, d_l\leq d_{ij}<d_m)}
    \end{split}
\end{equation}

The local $\hat{\tau}_i$ functions (Equation \ref{eq:lohandstein}) computed for the MPSB are similar to an application of a spatial bootstrap to the $K$-function \citep{Baddeley2015}, which like $\tau$ is a second-order correlation function. However we do \emph{not }recommend this literal interpretation of Loh \& Stein's method of averaging localised $\tau$-functions for the tau statistic, as the MMPSB method explains (\S\ref{S:bstrapmethod} \& \ref{S:MMBPadscaveats}), but provide it for completeness (Equation \ref{eq:loh6}).
\begin{equation}
\label{eq:loh6}
    \tau_{\textnormal{MPSB}}^*(d_l,d_m) = \frac{1}{n}\mathlarger{\mathlarger{\mathlarger{\sum_{i^*}}}}\frac{\theta_{i^*}(d_l,d_m)}{\theta_{i^*}(0,\infty)} = \frac{1}{n}\mathlarger{\mathlarger{\mathlarger{\sum_{i^*}}}}\frac{\Big(\frac{m_{i*}(d_l,d_m,k=1)}{m_{i*}(d_l,d_m,k=0)}\Big)}{\Big(\frac{m_{i*}(k=1)}{m_{i*}(k=0)}\Big)}
\end{equation}

\subsection{Advantages and caveats of the modified marked point spatial bootstrap (MMPSB)}
\label{S:MMBPadscaveats}
The schema (Equations A.\ref{eq:loh1}-\ref{eq:loh5}) is more robust than the original Loh \& Stein method (Fig.\ \ref{fig:loh}) when cases $i^*$ have no time-unrelated cases to pair with in their local distance band, i.e.\ $m_{i^*}(d_l,d_m,k=0) = 0$ in Equation \ref{eq:loh6} causes infinite values for $\theta_{i*}(d_l,d_m)$, or \texttt{NaN} values when also $m_{i^*}(d_l,d_m,k=1) = 0$; the MMPSB simply characterises these null events as zeroes and their addition in Equations \ref{eq:loh2} \& \ref{eq:loh3} separately protects the rest of the calculation. Alternative remedies to Loh \& Stein's approach such as dropping these contributions or merging contiguous distance bands that we attempted proved fruitless---the envelope diverged greatly for short distances and was biased above for larger distances and only 77\text{\textperiodcentered}3\% of simulations contributed to the CI compared to 100\% for MMPSB (Fig.\ \ref{fig:loh}). Dropping these inconvenient $i^*$ cases removes important spatial information which the tau bootstrap estimator in Equation \ref{eq:loh6} is sensitive to. 

Our method solves the numerical challenges but is not exactly the Loh \& Stein method as we indirectly obtain the tau estimate via calculation of the spatially bootstrapped odds $\theta^*$, so it is unclear if the validation of their results automatically transfers to our modified form. We also assume the mean of the bootstrap distribution of local mark functions asymptotically approximates the (global) tau statistic, as Loh \& Stein only provided experimental evidence to support this \citep{LohStein2004,Loh2008}.
\newpage
\FloatBarrier

\section{Additional figures}
\setcounter{figure}{0}
\begin{figure}[H]
\centering\includegraphics[width=0.99\linewidth]{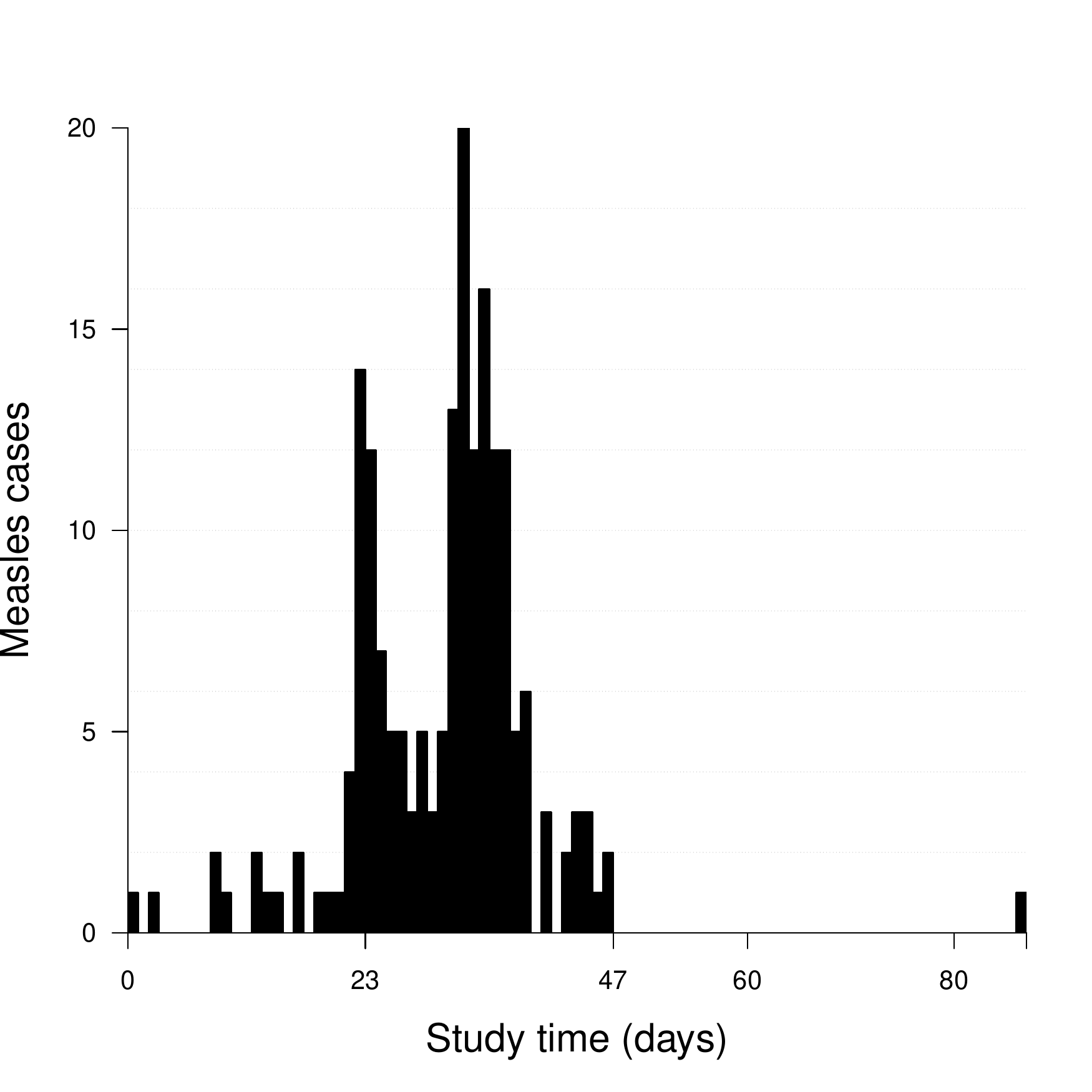}
\caption{Epidemic curve of the 188 measles cases in Hagelloch in 1861.}
\label{fig:re}
\end{figure}

\begin{figure}[H]
\centering\includegraphics[width=1.0\linewidth]{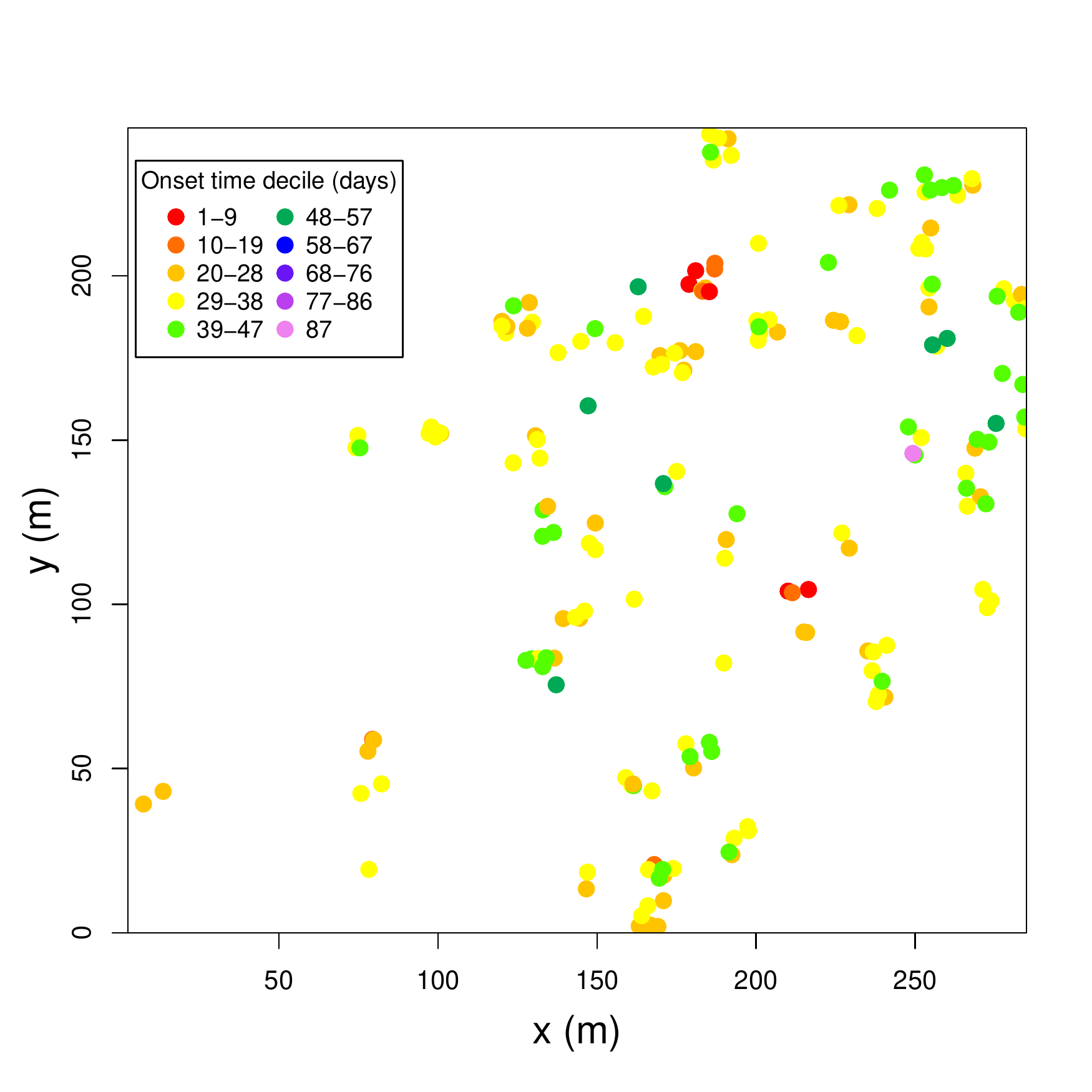}
\caption{Spacetime points of cases' locations with onset times as colour marks. Cases jittered up to 5m separately in $x$ and $y$ dimensions using the Uniform distribution to show multiple case households. There is some indication of cases in nearby households ($\sim$50m apart) having a similar date of onset, which may indicate direct transmission up to this distance.}
\label{fig:stplot}
\end{figure}

\begin{figure}[H]
\centering\includegraphics[width=1\linewidth]{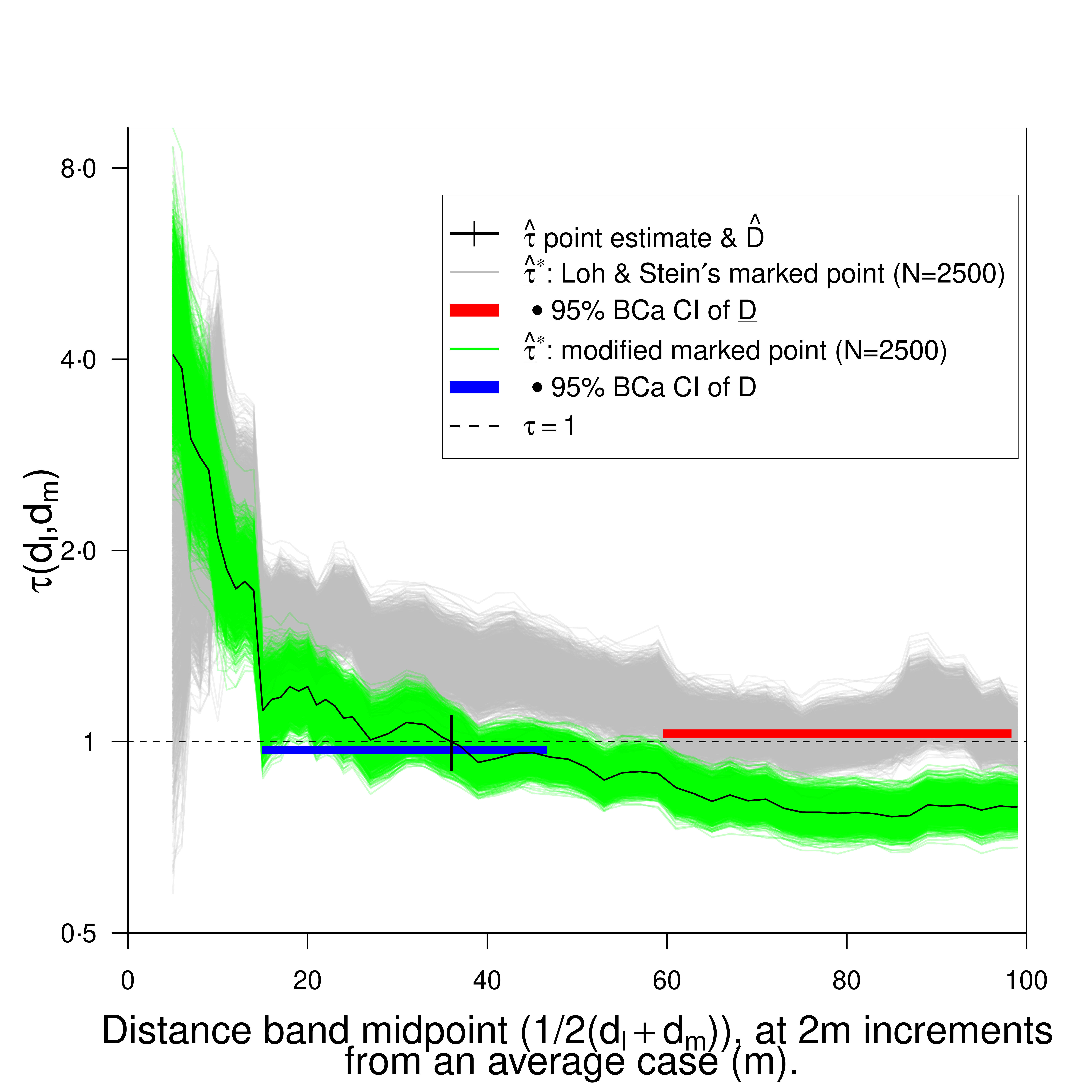}
\caption{MMPSB sampling compared with the original Loh \& Stein MPSB for the tau statistic. The latter's envelope $\hat{\underline{\tau}}^*$ poorly covers $\hat{\tau}$ at short distances and leads to over-bias in $\hat{\tau}$ at large distances; note that only 77\text{\textperiodcentered}3\% of tau spatial bootstrap simulations $\hat{\underline{\tau}}^*$ contribute to the MPSB BCa CI compared to 100\% for MMPSB. Distance band set as Fig.\ \ref{fig:get}, $N=$2500.} 
\label{fig:loh}
\end{figure}
\newpage
\FloatBarrier
\end{document}